\documentclass[11pt]{article}
\begin{document}
\renewcommand{\theequation}{\arabic{section}.\arabic{equation}}
\def\be{\begin{equation}}
\def\ee{\end{equation}}
\def\ba{\begin{eqnarray}}
\def\ea{\end{eqnarray}}
\def\nn{\nonumber}
\def\lb{\label}
\def\bb{\bibitem}
\def\dfrac{\displaystyle\frac}
\def\m{\overline\mu}
\def\S{\overline\Sigma}
\def\T{\overline\Theta}
\def\v{\overline{v}}
\def\p{\hat\varphi}
\def\U{\overline{U}}
\def\V{\overline{V}}
\def\W{\overline{W}}
\def\E{{\cal E}}
\begin{titlepage}

\date{3 July 2018}

\title{
\begin{flushright}\begin{small}    LAPTH-020/18
\end{small} \end{flushright} \vspace{1cm}
Rotating magnetized black diholes}

\author{G\'erard Cl\'ement\thanks{Email: gclement@lapth.cnrs.fr} \\ \\
{\small LAPTh, Universit\'e Savoie Mont Blanc, CNRS,} \\ {\small 9 chemin de Bellevue,
BP 110, F-74941 Annecy-le-Vieux cedex, France}}

\maketitle

\begin{abstract}
We analyze a four-parameter class of asymptotically flat magnetized solutions to the Einstein-Maxwell 
equations constructed by Manko et al., and show that these represent systems of two co-rotating extreme 
black holes with equal masses and electric charges, and opposite magnetic and NUT charges, connected 
by a cosmic string. We discuss several three-parameter subclasses, and determine in each case the 
parameter domain in which the ring singularity is absent. We find a two-parameter subclass and 
a one-parameter subclass where the conical singularity is also absent in the horizon co-rotating frame.
\end{abstract}
\end{titlepage}
\setcounter{page}{2}

\section{Introduction}
Since the early days of general relativity, it has been known that
axisymmetric multi-black hole solutions to the Einstein equations can
be constructed from linear superpositions of one-black hole solutions
\cite{bachweyl}. These generically present conical singularities
\cite{israelkhan} on the symmetry axis (cosmic strings) to account for
the forces necessary to balance the attraction between black holes, the
only known exception being the static Majumdar-Papapetrou \cite{papa} linear
superposition of identical extreme Reissner-Nordstr\"om black holes.

The existence of more general, non-linear multi-black hole solutions
to the Einstein or Einstein-Maxwell equations has been proved in the
harmonic map analysis of \cite{weinstein96}. A number of such rotating
solutions have been constructed by inverse scattering techniques, see
the review in \cite{GP}, the most recent work on the subject being probably
\cite{2kerr} on the two-Kerr system. However little is generally known about their
structure. It was first shown by Emparan \cite{emparan} that the static
magnetized Bonnor solution \cite{bonnor66} actually represents a black
dihole, a system of two black holes with opposite magnetic charges and
degenerate horizons, held apart by a cosmic string. This string can be
removed by applying an external magnetic field, at the expense of
asymptotic flatness. Recently, we have shown \cite{tale} that a rotating
solution to the Einstein-Maxwell equations previously constructed in \cite{GC98}
represents a more complex system of two extreme black holes with equal
masses and electric charges, and opposite magnetic and gravimagnetic
(NUT) charges, co-rotating at an angular velocity fine-tuned to one-fourth
of the inverse NUT charge. A third, necessary partner in this system is
an electrically charged, magnetized strut or string which also acts as
a Dirac-Misner string.

Both the Bonnor solution and the solution studied in \cite{tale} belong to
a larger four-parameter class of solutions constructed with the aid of
Sibgatullin's method \cite{sibga84} by Manko et al. in \cite{manko00a},
where their connection with neutron star models was suggested. Perhaps 
because of their complexity, the properties of these solutions have not, 
to our knowledge, been investigated. We will
show in the present paper that all these solutions also represent co-rotating
magnetized black diholes, with characteristics similar to those elucidated in
\cite{tale}: equal masses and electric charges, opposite magnetic and NUT
charges, and the existence of a cosmic-Dirac-Misner string connecting the
two degenerate horizons. For several three-parameter subclasses, we will
discuss the algebraic conditions for the absence of ring singularities -- a
necessary condition for the global solution to be quasi-regular, with only a mild
conical singularity. Interestingly, we shall also find a two-parameter
subclass and a one-parameter subclass where the conical singularity itself
is absent in the horizon co-rotating frame, leading to apparently regular 
two-black-hole solutions.

The solutions of \cite{manko00a} are presented in the next section. In section 3
we discuss the properties of the horizons and compute the various horizon charges.
The interconnecting string is considered in section 4. Several three-parameter
subclasses are further discussed in the next three sections, and our conclusions
are presented in the last section.

\setcounter{equation}{0}
\section{The solutions}
\subsection{General form} 
Let us start from Manko et al's four-parameter asymptotically
flat rotating magnetized solution \cite{manko00a} (a subfamily of the nine-parameter
electrovac solutions of \cite{manko00b}). In a first step, we will
choose as independent parameters the overall scale $\kappa>0$, and three
dimensionless parameters $m>0$, $a$ and $b$ which are those of Manko
et al. (here indexed with $M$) divided by $\kappa$ (noted $k$ in \cite{manko00a}), i.e.
$m = M_M/\kappa$, $a = a_M/\kappa$, $b = b_M/\kappa$. Auxiliary
parameters are $d = d_M/\kappa^2$, $\delta = \delta_M/\kappa^2$
and $\m = \mu_M/\kappa^2$ defined by
 \be\lb{ddeltamu}
4d=m^2-(a-b)^2, \quad \delta = 1-d, \quad \m^2 = m^2b^2+4d\delta.
 \ee
The physical parameters, total mass $M$, total angular momentum $J$
and total dipole magnetic moment $\mu$ are related to the preceding by
 \be\lb{paras}
M = \kappa m, \quad J = \kappa^2 ma, \quad \mu = \kappa^2\m.
 \ee

The Ernst potentials\footnote{We use the same conventions for defining the Ernst
potentials as in \cite{smarr}, equations (3.1)-(3.3).}
$\E$, $\psi$ may be expressed in terms of
Kinnersley potentials $(U,V,W)$ according to
 \be (1970)
{\cal E} = (U-W)/(U+W), \qquad \psi = V/(U+W),
 \ee
The Kinnersley potentials for the Manko et al. family of solutions
(labelled $A^\ast/4\kappa^4,i\mu C^\ast/2\kappa^4,mB^\ast/2\kappa^4$ in \cite{manko00a},
where $*$ denotes the complex conjugate) are given in prolate spheroidal coordinates by:
 \ba\lb{kin}
U &=& (x^2-\delta y^2)^2 - d^2 - \nu\lambda(1-y^4)
+ 2ixy[\nu(x^2-1)+\lambda(1-y^2)], \nn\\
V &=& \m\{-\nu x(1-y^2) + iy[(x^2-1)+\delta(1-y^2)]\}, \nn\\
W &=& mx[(x^2-1)+(b\nu+\delta)(1-y^2)]- \nn\\ && -
imy[b(x^2-1)+(b\delta-\lambda)(1-y^2)].
 \ea
The prolate spheroidal coordinates $x\ge1$, $y\in[-1,+1]$ are related to
the Weyl cylindrical coordinates $\rho$, $z$ by
 \be\lb{sphero}
\rho = \kappa\sqrt{(x^2-1)(1-y^2)}, \quad z = \kappa xy.
 \ee

To simplify the form of the solution, we have introduced two new dimensionless parameters 
$\nu$ and $\lambda$ which play symmetrical roles in (\ref{kin}). These are related to our original parameters by
 \be
\nu \equiv (a-b)/2, \quad \lambda \equiv \nu(d-\delta) - m^2b/2.
 \ee
The other dimensionless parameters occurring in (\ref{kin}) are related to $m$, $\nu$ and $\lambda$ by
 \be
\delta = 1 + \nu^2 - \frac{m^2}4, \quad d = 1-\delta, \quad b = \nu - \frac4{m^2}\left[\nu^3 
+ \frac{\nu+\lambda}2\right],
 \ee
$\m$ being given in terms of these by the last equation (\ref{ddeltamu}) \footnote{Note that the reality of $\m$ 
is ensured only in a sector of the three-space $(m,\nu,\lambda)$.}.  
We note also that, from (\ref{kin}), the total quadrupole electric moment is $Q_2 = -\kappa^3\m\nu$, so that 
$\nu$ is a measure of the electric quadrupole to magnetic dipole ratio.

The corresponding metric and electromagnetic potential may be written in the generic form
 \ba\lb{ansatz}
ds^2 &=& - \frac{f}\Sigma\left(dt-\frac{\kappa\Pi}f d\varphi\right)^2 + \nn\\
&+& \kappa^2\Sigma\left[(x^2-y^2)^{-3}\left(\frac{dx^2}{x^2-1} +
\frac{dy^2}{1-y^2}\right) + f^{-1}(x^2-1)(1-y^2)d\varphi^2\right], \nn\\
A &=& \frac{1}{\Sigma}[\v dt + \kappa\Theta d\varphi],
 \ea
where the various functions, evaluated in \cite{manko00a}, are
 \ba\lb{sol}
&f(x,y) &= [\zeta^2 + \nu\lambda(1-y^2)^2]^2 - 4(x^2-1)(1-y^2)(\nu x^2-\lambda y^2)^2, \nn\\
&\Sigma(x,y) &= \left\{\zeta(\zeta+mx+2d) + mb\nu x(1-y^2) - \lambda\nu(1-y^4)\right\}^2 + \nn\\
&& + y^2\left\{2x[\nu(x^2-1)+\lambda(1-y^2)] + m[-b\zeta + \lambda(1-y^2)]\right\}^2, \nn\\
&\Pi(x,y) &= - (1-y^2)\left\{(x^2-1)(\nu x^2-\lambda y^2)
\left(4mx[\zeta+mx+2d-b\nu(1+y^2)] + \right.\right. \nn\\ && \left. + 2(m^2b^2-4d\delta)y^2\right)
+ [\zeta^2+\nu\lambda(1-y^2)^2]\cdot \nn\\ &&\cdot \left.\left(2mb(x+m)\zeta + [-m\lambda(2x+m) +
\frac\nu2(m^2b^2-4d\delta)](1-y^2)\right)\right\}, \nn\\
&\v(x,y) &= \m\left\{-\nu x(1-y^2)\left(\zeta(\zeta+mx+2d) + mb\nu x(1-y^2) - \lambda\nu(1-y^4)\right)\right. +\nn\\
&& \left. + y^2\zeta\left(2x[\nu(x^2-1)+\lambda(1-y^2)] + m[-b\zeta + \lambda(1-y^2)]\right)\right\} \nn\\
&\Theta(x,y) &= \frac{\m(1-y^2)}2\left\{\left[\zeta(\zeta+mx+2d) + mb\nu x(1-y^2)
- \lambda\nu(1-y^4)\right]\cdot \right. \nn\\ && \cdot \left[(2x+m)(\zeta + m^2) + 2m(x^2-d-2\nu^2))
- mb\nu(1+y^2)\right] + \nn\\
&& + 2y^2\left[2x(\nu(x^2-1)+\lambda(1-y^2)) + m(-b\zeta + \lambda(1-y^2))\right] \cdot \nn\\
&& \left. \cdot \left[mb(x+m)-\nu x^2 + \lambda)\right]\right\}.
\ea
In the preceding, we have put
 \be
\zeta \equiv x^2-1+\delta(1-y^2).
 \ee
The angular variable $\varphi$ is assumed to have the standard periodicity $2\pi$, so that 
the metric is asymptotically flat.

In the special case $\delta=\lambda=0$, we recover the special rotating dihole solution 
analyzed in \cite{tale}. We shall show in the following that the more
general solution discussed here shares similar properties. The metric (\ref{ansatz}) describes a system
of two co-rotating black holes of horizons $x=1$, $y=\pm1$ (discussed in section 3), connected by a string 
(a segment along which the metric has a conical singularity) $x=1$, $y^2<1$ (discused in section 4). 
There is also an ergosphere, where $f(x,y)<0$, generically
bounded by two surfaces connecting the two ends of the string ($f$ is positive on
the string, except if $\delta^2+\nu\lambda=0$). The two black holes generically have equal
masses and electric charges, and opposite NUT charges and magnetic charges. This metric can also present 
a naked ring singularity, which is the locus
where the function $U+W$ vanishes. As opposed to the conical singularity of the string, this is a strong
curvature singularity, so only the solutions where this ring singularity is absent can be considered as
quasi-regular. 

\subsection{Special cases}
Although with our parameterization the general form (\ref{kin}) of the Kinnersley potentials 
looks rather simple, the discussion of the physical properties turns out to be very intricate 
in the general four-parameter case, due to the non-linear dependence of the auxiliary parameters 
$\delta$, $d$, $b$, and $\m$ on the dimensionless parameters $m$, $\nu$ and $\lambda$, and the 
additional constraint on the reality of $\m$. Especially, the absence of ring singularity is 
technically difficult to prove (or disprove) in the general case. In the present paper, we will only 
discuss this question in the case of several special three- or two-parameter sub-spaces.

a) \underline{$\delta=\lambda=0$}. This class of solutions depending only on two parameters, 
e.g. the total mass and the total angular momentum, is a magnetized rotating generalization of 
the ``$\delta=2$'' \footnote{This parameter $\delta$ has no relation with the parameter $\delta$ 
used in the present paper.}
static Zipoy-Voorhees \cite{ZV} vacuum solution, different from the ``$\delta=2$'' rotating 
Tomimatsu-Sato \cite{TS} vacuum solution (TS2). These solutions were discussed in detail in \cite{tale},
the parameters introduced in the present paper being related to those of \cite{tale} by
 \be\lb{parastale}
m=\frac2p, \quad \nu  = \frac{q}p, \quad b = \frac{pq}{2}, \quad \m = -\varepsilon q
 \ee
($q^2+p^2=1$). Remarkably, all the solutions of this class are free from a naked ring singularity. 
We will not discuss further these solutions here. 
 
b) \underline{$2m+d=0$}. The solutions of this three-parameter class, analyzed in section 5, have
electrically neutral constituents (horizons and string). We will show that the ring singularity is 
absent in two disjoint three-parameter subsectors.

c) \underline{$\nu=0$}. This constraint corresponds to the vanishing of the total electric quadrupole
moment $Q_2$, resulting from a delicate balance between electrically charged horizons and string. 
This three-parameter class is discussed in section 6, where we show that the ring singularity is 
absent in the sector $m<2$. We also single out two two-parameter subclasses. The first subclass 
$b=0$ ($\lambda=0$) coincides with the static Bonnor solution \cite{bonnor66}, while the second
subclass $b=b_c(m)$ (given in (\ref{bc})) is characterized by a vanishing string tension in the 
horizon co-rotating frame.

d) \underline{$\lambda=\nu$}. This three-parameter class, characterized by a vanishing horizon 
angular velocity is discussed in section 7, where we identify a ring singularity-free sector 
(equation (\ref{regstat})). Three special two-parameter subclasses are $\nu=0$ (the Bonnor solution),
$\delta=0$ (the TS2 solution), and $\delta=m^2/2$ ($\nu^2=3m^2/4-1$), leading to a vanishing 
total angular momentum. This last subclass contains a one-parameter family $m=m_c\simeq1.30$ with 
again a vanishing string tension.

Before analyzing these various special cases, we discuss in the next two sections the general physical 
properties of the two-component horizon and of the interconnecting string. 

\setcounter{equation}{0}
\section{Horizons}

The ``points'' $x=1$, $y=\pm1$ ($\rho=0$, $z=\pm\kappa$) are actually horizons, two-surfaces with a finite area
which shall be computed below. To see this, take the limit $x\to1$ and $y\to\pm1$ with the ratio
 \be\lb{X2}
X^2 = \frac{1-y^2}{x^2-1}
 \ee
held fixed. In this limit, the reduced (barred) Kinnersley potentials defined by
$(U,V,W)=(\U,\V,\W)(x^2-1)$ go to
 \ba\lb{kinh}
\U &=& 2\{d + (d\delta-\nu\lambda)X^2 \pm i(\nu+\lambda X^2)\}, \nn\\
\V &=& \m\{\nu X^2 \pm i(1+\delta X^2)\}, \nn\\
\W &=& m\{1 + (b\nu+\delta)X^2 \mp i[b+(b\delta-\lambda)X^2]\}.
 \ea
In the same limit, the auxiliary function $\zeta(x,y)$ behaves as
$\zeta \sim [1+\delta X^2](x^2-1)$, leading for $\lambda\neq\nu$ to
 \ba
f &\sim& -4(\nu-\lambda)^2X^2(x^2-1)^2, \nn\\
\Sigma &\sim& \S(X)(x^2-1)^2, \nn\\
\Pi &\sim& 2(\lambda-\nu)\left[(mb-2\nu)^2 + (m+2d)^2\right]X^2(x^2-1)^2,
 \ea
where
 \ba
\S(X) &=& \left(m+2d + \left[(m+2d)\delta + (mb-2\lambda)\nu\right]X^2\right)^2 + \nn\\
&+& \left(2\nu-mb + \left[(m+2d)\lambda - (mb-2\lambda)\delta\right]X^2\right)^2.
 \ea

It follows that
 \be\lb{gpp}
g_{\varphi\varphi} = \kappa^2\left[\frac{\Sigma}f(x^2-1)(1-y^2) -
\frac{\Pi^2}{\Sigma f}\right] \sim -\frac{\kappa^2\Pi^2}{\Sigma f}
 \ee
is finite and positive, while the lapse
 \be
N^2 = \frac{\kappa^2(x^2-1)(1-y^2)}{g_{\varphi\varphi}}
 \ee
develops a double zero at $x=1$, $y=\pm1$, corresponding to two
double horizons $H_\pm$, co-rotating at the angular velocity
 \be\lb{OmH}
\Omega_H = \left. \frac{f}{\kappa\Pi}\right\vert_H =
\frac{2(\nu-\lambda)}{\kappa[(m+2d)^2 + (mb-2\nu)^2]}.
 \ee

Let us transform from $(x,y)$ to the coordinates \cite{KoHi} $X$ and $Y = y/x$. Noting
that, near the horizons $Y = \pm1$,
 \be
-\frac{\Pi^2}f \sim \S_0^2X^2(x^2-1)^2,
 \ee
where $\S_0 \equiv \S(0)$, one finds that the horizon metric
degenerates to
 \be\lb{methor}
ds_H^2  = \kappa^2\S(X)\left[\frac{dX^2}{(X^2+1)^4} +
\left(\frac{\S_0}{\S(X)}\right)^2X^2\,d\p^2\right],
 \ee
in the co-rotating near-horizon frame $(\hat{t},X,Y,\p)$ defined by
$\hat{t} = t$, $\p = \varphi - \Omega_H t$. The horizon area is
 \be\lb{area}
{\cal A}_H = \pi\kappa^2\S_0 = \pi\kappa^2\left[(m+2d)^2+(mb-2\nu)^2\right].
 \ee
The horizon metric
(\ref{methor}) is regular at $X=0$ ($y=\pm1$), but generically presents a
conical singularity at $X\to\infty$ ($x=1$) with deficit angle
$2\pi(1-\alpha_H)$, where
 \be\lb{alphaH}
\alpha_H = \frac{\S_0}{\S_4},
 \ee
with
 \be
\S_4 = [(m+2d)\delta + (mb-2\lambda)\nu]^2 + [(m+2d)\lambda - (mb-2\lambda)\delta]^2
 \ee
the coefficient of $X^4$ in $\S(X)$. Again, it is difficult to determinate the sign of
this deficit angle in the general case.

The electromagnetic potential on the horizon is, in the co-rotating frame,
 \ba
\hat{A} &=& - \frac{\m(mb-2\nu)}{[(mb-2\nu)^2+ (m+2d)^2]}\,dt + \frac{\kappa\T(X)}{\S(X)}\,d\p, \\
\T &=& \frac\m2X^2\left\{\left((m+2)(m+2d)-2\nu(mb-2\nu)\right)\left(m+2d +
[(m+2d)\delta + (mb-2\lambda)\nu]X^2\right) + \right.\nn\\&& \left. + 2\left(mb(m+1)+\lambda-\nu\right)
\left(2\nu-mb + [(m+2d)\lambda - (mb-2\lambda)\delta]X^2\right)\right\}. \nn
 \ea
The horizon vector potential generates a magnetic field perpendicular to the horizon,
flux conservation implying that the two horizons carry opposite magnetic charges
$\pm P_H$, with
 \be\lb{P}
\left. P_H=-\frac1{4\pi}\oint_{H_+}dA_{\varphi}\,d\varphi = -\frac12A_{\varphi}(X)\right\vert_\infty^0 = \frac{\kappa\T_4}{2\S_4},
 \ee
where $\T_4$ is the coefficient of $X^4$ in $\T(X)$.

Computation of the horizon electric charges, which are equal because the electric potential is even in $y$,
necessitates in principle the knowledge of
the electric field off the horizon. However we can use for this purpose the Tomimatsu formula
\cite{tom84}, which necessitates only the knowledge of the Kinnersley potentials on the horizon
and that of the horizon angular velocity:
 \be\lb{QH}
\left. Q_H = \dfrac1{4\pi}\oint_{H_+}\omega\,\dfrac{d\,{\rm
Im}\psi}{dX}\,dXd\varphi = \dfrac{\omega_H}2{\rm
Im}\psi(X)\right\vert_\infty^0
 \ee
with $\omega_H=1/\Omega_H$ the constant value of $\omega = \kappa\Pi/f$ over the horizon, and
 \be\lb{impsi}
{\rm Im}\psi = \frac{{\rm Re}(\U+\W){\rm Im}\V - {\rm Re}\V{\rm
Im}(\U+\W)}{|\U+\W|^2}.
 \ee
The result is
 \be\lb{QH1}
Q_H = \frac{\kappa\m(m+2d)}{4\S_4}\left[\nu m^2b^2 - 4\nu(\delta^2+\nu\lambda) + 2\delta mb(m+2) - \lambda(m+2)^2\right].
 \ee

The horizon Komar mass and angular momentum are given by the Tomimatsu formulas \cite{tom84}
(as corrected in \cite{smarr}), written here for degenerate black holes:
 \ba
M_H &=& \frac1{8\pi}\oint_H\left[\omega\,d\,{\rm Im}\E +
2d(A_\varphi\,{\rm Im}\psi)\right]d\varphi,\lb{MH}\\
J_H &=& \frac1{8\pi}\oint_H\omega\left[\dfrac12\omega\,d\,{\rm Im}\E
+ d(A_\varphi\,{\rm Im}\psi) + \omega\hat{A}_t\,d\,{\rm
Im}\psi\right]d\varphi. \lb{JH}
 \ea
The computation gives, after some simplifications,
 \ba\lb{MH1}
M_H &=& \frac{\kappa m}2\frac1{\S_4}\left[(\delta^2-bd\lambda)(m+2d)^2+ 2\delta\gamma(bd+\nu)(m+2d) + \right.\nn\\
&& \left. + 2bd\delta^2(\beta+\gamma) + (\delta^2+bd\nu)(m^2b^2-4\lambda\nu) + mb\nu(\beta\lambda+\gamma\nu)
\right] -\nn\\
&& - \frac{\m(m+2d)(\delta^2+\nu\lambda)P_H}{\S_4},
 \ea
where we have put
$$\beta \equiv mb-2\nu, \quad \gamma \equiv mb-2\lambda.$$
The second term is the contribution of the second piece of
(\ref{MH}), which was missing in the original Tomimatsu formula. This term, proportional to the
product of the horizon electric and magnetic charges, vanishes for $m+2d=0$, as well as for
$\delta=\lambda=0$ (the case studied in \cite{tale}). As shown in \cite{smarr}, the system of
(\ref{JH}), (\ref{MH}) and (\ref{QH}) leads to the horizon Smarr formula for degenerate
black holes
 \be\lb{smarr}
M_H = 2\Omega_HJ_H + \Phi_HQ_H,
 \ee
with $-\Phi_H = \hat{A}_t$ the horizon electric potential in the co-rotating frame.

\setcounter{equation}{0}
\section{String}
The string is the coordinate singularity $x=1$, $y^2<1$ ($\rho=0$, $-\kappa<z<\kappa$).
Near the string, i.e. for $\xi^2 \equiv x^2-1\to 0$, using $\zeta=\delta(1-y^2)+\xi^2$,
we obtain
 \ba
f &\simeq& f_S = (\delta^2+\nu\lambda)^2(1-y^2)^4 \quad {\rm if} \quad
\delta^2+\nu\lambda \neq 0, \\
f &\simeq& f_C = -4(1-y^2)(\nu-\lambda y^2)^2\xi^2 \quad {\rm if} \quad
\delta^2+\nu\lambda = 0.
 \ea
This second case is that of a spinning cosmic string (in a background
curved spacetime), with spin $\Omega_H^{-1}/4$, rotating at the critical
angular velocity $\Omega_H$. We will not treat here this case, similar
to that considered in \cite{tale}, and discuss only the case
$\delta^2+\nu\lambda\neq0$.

In this case,
 \ba
\Sigma_S &\simeq& \left\{[(mb-2\lambda)\nu + (m+2d)\delta + (\delta^2+\nu\lambda)(1-y^2)]^2 + \right. \nn\\
&& \left. + [(mb-2\lambda)\delta-(m+2d)\lambda]^2y^2\right\}(1-y^2)^2,\\
\Pi_S &\simeq& - (\delta^2+\nu\lambda)\left[2mb(m+1)\delta-m(m+2)\lambda
+\dfrac\nu2(m^2b^2-4d\delta)\right](1-y^2)^4, \nn
 \ea
and the near-string metric is
 \be\lb{string}
ds^2 \simeq -F_S(dt-\omega_Sd\varphi)^2 + \frac{\kappa^2\Sigma_S}{(1-y^2)^3}\left[
\frac{dy^2}{1-y^2}+d\xi^2+\alpha_S^2 \xi^2 d\varphi^2\right],
 \ee
where $F_S=f_S/\Sigma_S$, $\omega_S=\kappa\Pi_S/f_S$ is a constant, and
 \be
\alpha_S = \frac1{\delta^2+\nu\lambda}.
 \ee

What is relevant for comparison with the horizon metric (\ref{methor}) is
the near-string metric in the horizon co-rotating frame ($d\varphi=d\hat\varphi
+ \Omega_Hdt$),
 \be\lb{string1}
ds^2 \simeq -\hat{F}_S(dt-\hat\omega_Sd\p)^2 + \frac{\kappa^2\Sigma_S}{(1-y^2)^3}\left[
\frac{dy^2}{1-y^2}+d\xi^2+\hat\alpha_S^2 \xi^2 d\p^2\right],
 \ee
where
 \be\lb{stringpara}
\hat{F}_S = (1-\Omega_H\omega_S)^2\,F, \quad \hat\omega_S = \frac{\omega_S}
{(1-\Omega_H\omega_S)}, \quad  \hat{\alpha}_S = \frac{\alpha_S}{(1-\Omega_H\omega_S)}.
 \ee
A lengthy computation yields, after simplification,
 \be
\hat{\alpha}_S = \alpha_H,
 \ee
where $\alpha_H$ is given by (\ref{alphaH}), i.e. the string and horizon
conical singularities are equal, when computed in the same reference frame.

As shown in \cite{tale}, the spin of the spinning cosmic string
metric (\ref{string1}) should actually be interpreted
as a gravimagnetic flow along the Misner string connecting two NUT sources
at $\rho=0$, $z = \pm\kappa$, with NUT charges $\pm N_H$,
 \be\lb{nut}
N_H = - \frac{\hat\omega_S}4 = \frac{\alpha_S-\alpha_H}{4\Omega_H\alpha_S}.
 \ee
One obtains
 \be\lb{N}
N_H = \frac{\kappa\alpha_H}8\left[4\delta^2(mb-\lambda-\nu)
+ \nu(m^2b^2-4\lambda\nu)+2\delta(m+2d)(mb-2\lambda)-\lambda(m+2d)^2\right].
 \ee
This value, as well as those of the other black hole observables, reduce
to those given in \cite{tale} for $\delta=\lambda=0$.

\setcounter{equation}{0}
\section{The neutral class of solutions}

Equation (\ref{QH1}) shows that the horizon electric charge vanishes for
 \be\lb{cons1}
m+2d = 0.
 \ee
Because the global electric charge has been set to zero, the string electric
charge will also vanish. We will refer to the class ($m+2d=0$) of solutions with
electrically neutral constituents as the neutral class.

The constraint (\ref{cons1}) leaves only the two independent dimensionless parameters $m$ and
$b$, the other parameters being related to these by
 \be\lb{rel1}
\delta = \frac{m+2}2,\; \nu^2 = \frac{m(m+2)}4,\; \lambda =
-\left[(m+1)\nu + \frac{m^2b}2\right],\; \m^2 = m^2b^2 - 4\nu^2,
 \ee
this last relation implying the restriction
 \be\lb{bound1}
b^2 > \frac{m+2}m.
 \ee

\subsection{Absence of ring singularity}
We wish to further constrain the parameters in this class so that the ring singularity is absent.
The ring singularity corresponds to a zero of $\Sigma(x,y)$, i.e. to
a solution of the system
 \be\lb{ring}
{\rm Re}(U+W) = 0, \qquad {\rm Im}(U+W) = 0
 \ee
with $y^2\neq1$. The first equation
(\ref{ring}) reads, for $m+2d = 0$,
 \be
\zeta[\zeta+m(x-1)]  + (m+1)\nu^2(1-y^2) + mb\nu\left[x+\frac{m(1+y^2)}2\right](1-y^2) = 0.
 \ee
With $x\ge1$, $y^2<1$ and $m>0$, $\delta>0$, a {\em sufficient} condition for the absence of solutions to this
equation, and thus absence of ring singularity, is obviously $b\nu\ge0$,
implying on account of (\ref{rel1}) and (\ref{bound1}) $b\nu>(m+2)/2$, or $mb/2\nu>1$.

The second equation (\ref{ring}) is trivially satisfied in the
equatorial plane $y=0$, in which case the first equation reduces to
the quartic equation in $x$:
 \be
x^4 + mx^3 + m(b\nu-d)x - (d^2+\nu\lambda) = 0.
 \ee
It is clear that for $b\nu<0$ and large enough this equation
will have a solution for some $x>1$. For $x=1$, $m+2d = 0$, the equation reduces to
 \be
2mb\nu + m(m+2)+ 2 = 0,
 \ee
so that a {\em necessary} condition for the equatorial ring singularity to be absent is
 \be\lb{range2}
2mb\nu + m(m+2) > -2, \quad {\rm or} \quad \frac{mb}{2\nu} > -1 - \frac2{m(m+2)}.
 \ee
If $y\neq 0$, $1-y^2$ can be eliminated between the two equations
(\ref{ring}), leading to an equation of sixth degree in $x$. It is
not clear whether this equation can have a solution for $mb/2\nu<-1$, and
in that case whether such solutions (non-equatorial ring
singularities) could be excluded for some range of parameter values.

To conclude this subsection, the neutral solution is certainly free from a ring singularity
in the sector $mb/2\nu>1$, and possibly free from a ring singularity in the distinct sector
$-1-2/m(m+2) < mb/2\nu < -1$. We will see in the following that the properties of the two-black
hole system in these two sectors are quite different.

\subsection{Black hole properties}
The values of the various horizon parameters computed in the general case simplify in this case to
 \ba\lb{neutchar}
\Omega_H &=& \frac{2(\nu-\lambda)}{\kappa(mb-2\nu)^2}, \quad
\alpha_H = \frac1{\delta^2+\nu^2}\left(\frac{mb-2\nu}{mb-2\lambda}\right)^2, \nn\\
M_H &=& - \frac{\kappa m(mb-2\nu)(\delta^2+mb\nu/2)}{2(mb-2\lambda)(\delta^2+\nu^2)},  \nn\\
N_H &=& \frac{\kappa\alpha_H}8\left[4(\delta^2+\nu^2)(mb-\lambda-\nu)
+ \nu(mb-2\nu)^2\right], \\
Q_H &=& 0, \quad P_H = - \frac{\kappa\m(mb-2\nu)}{2(mb-2\lambda)},
 \ea
and the horizon angular momentum
 \be\lb{MJH}
J_H = \frac{M_H}{2\Omega_H}
 \ee
from the Smarr formula for an electrically neutral degenerate black hole.

It follows from the last equation (\ref{rel1}) that, for a vacuum solution ($\m=0$), either one of the two
combinations $(mb-2\nu)$ and
$$(mb-2\lambda) = (m+1)(mb+2\nu)$$
must vanish, with unpleasant consequences
(for $mb-2\nu=0$ the horizon area (\ref{area}) vanishes, while for $mb-2\lambda=0$, the angular deficit
and the horizon Komar mass diverge). On the other hand, for a non-vacuum solution ($\m\neq0$), the
combinations $(mb-2\nu)$ and $(mb-2\lambda)$ do not vanish and are of the same sign.
It then follows from the regularity condition (\ref{range2}), which implies $\delta^2+mb\nu/2 > (m+1)/2$,
that the horizon Komar-Tomimatsu mass $M_H$ is negative definite. While this conclusion seems surprising,
we should keep in mind the well-known fact that energy is not localisable in general relativity. As
demonstrated by Tomimatsu \cite{tom84} and discussed
in detail in \cite{smarr}, the total Komar mass, given by the integral over a spacelike surface at infinity
$$M = \frac1{4\pi}\oint_\infty D^\nu k^{\mu}d\Sigma_{\mu\nu}$$
(where $k^\mu = \delta^\mu_t$ is the Killing vector associated with time translations),
which is positive definite by the positive energy theorem, can be transformed in the Einstein-Maxwell case
into a sum $M = \sum_n M_n$ over spacelike surfaces $\Sigma_n$ bounding the various ``sources''. In the present
dihole case, this sum is
 \be\lb{totalmass}
M = 2M_H + M_S,
 \ee
where $M_H$ is given by (\ref{MH}), and
 \be\lb{MS}
M_S = \frac1{8\pi}\oint_{\Sigma_S}\left[g^{ij}g^{ta}\partial_jg_{ta}
+2(A_t F^{it}-A_\varphi F^{i\varphi})\right]d\Sigma_i,
 \ee
where $\Sigma_S$ is a small cylinder of radius $\xi$ surrounding the string $x=1$, $-1<y<1$, and
$x^a = (t,\varphi)$. The individual contributions to the sum (\ref{totalmass}) do not necessarily have a
definite sign (see e.g. the discussion in \cite{tom83}), only their sum $M=\kappa m$ must be positive.
In the present case, $M_H<0$ means that $M_S$ must be positive.

The necessary regularity condition (\ref{range2}) also implies that
 \be
\nu(\nu-\lambda) = \frac{m^2b\nu}2 + (m+2)\nu^2 > \frac{m(m+1)}2,
 \ee
which is positive definite. This means that the horizon angular velocity $\Omega_H$ never vanishes,
and has the same sign as $\nu$, while the sign of the Komar horizon angular momentum $J_H$ is opposite to
that of $\nu$. On the other hand, the {\em total} angular momentum $J=\kappa^2m(b+2\nu)$ does not have a definite
sign, and vanishes for $b=-2\nu$, the positivity of $\m^2$ together with
the necessary regularity condition (\ref{range2})
implying that the corresponding mass parameter $m$ must then lie in the range $1<m<\sqrt2$.

The parameter $\alpha_H$ associated with the conical singularity of the horizon and string can be reexpressed as
 \be
\alpha_H = \frac1{k^2}\,\left(\frac{mb-2\nu}{mb+2\nu}\right)^2,
 \ee
with
 \be
k^2 = \frac{(m+1)^3(m+2)}2 > 1.
 \ee
It follows that, in the first regularity sector $mb/2\nu>1$, $\alpha_H<1$, i.e. the string tension is positive.
The situation is different in the second sector $mb/2\nu<-1$ with the condition (\ref{range2}), which can be rewritten as
 \be
-\frac{mb}{2\nu} < \frac{(m+1)^2+1}{(m+1)^2-1} < \frac{k+1}{k-1},
 \ee
the last relation following from $k<(m+1)^2$. The conclusion is then that, in this second sector, $\alpha_H>1$, i.e. the
string tension is negative.

Expanding the bracket in the expression (\ref{neutchar}) for the horizon NUT charge gives
 \be
N_H = \frac{\kappa\alpha_H(m+2)\nu}8\left\{mz^2 + 2[(m+1)^2+1]z+m(2m+3)\right\},
 \ee
with $z= mb/2\nu$. The large bracket has two negative roots
 \be
z_\pm = \frac1m\left[-(m+1)^2-1\pm(m+1)\sqrt{m^2+4}\right],
 \ee
and is obviously positive definite in the first solution sector $z>1$, so that the sign of the
horizon NUT charge is equal to that of the parameter $\nu$, and opposite to that of the
horizon Komar angular momentum $J_H$. We can show that, in the second regular solution sector,
 \be
z_- < -1 - \frac2{m(m+2)} < z < -1 < z_+,
 \ee
so that the sign of the horizon NUT charge, and thus also of the NUT dipole $2\kappa
N_H$, is now equal to that of the horizon Komar angular momentum $J_H$.

\setcounter{equation}{0}
\section{The Bonnor class of solutions}
The analysis also simplifies when the total quadrupole electric moment vanishes,
 \be
\nu = 0,
 \ee
leading to
 \be\lb{parasB}
d=m^2/4, \quad \m^2 = m^2[1-m^2/4+b^2].
 \ee
Then the parameter $b$ coincides with the (scaled) angular momentum to mass ratio $a$.
If further $b=0$ we recover the Bonnor magnetostatic solution $a=b=0$ \cite{bonnor66},
so that we will refer to this class of solutions as the Bonnor class.

For the Bonnor class, the first equation (\ref{ring}) for the ring singularity reads
 \be
[x^2-1+\delta(1-y^2)][x^2-1+\delta(1-y^2) + mx +m^2/2] = 0,
 \ee
with
 \be
\delta = 1 - \frac{m^2}4.
 \ee
If $m>2$, the first bracket vanishes for $x=m/2$, $y=0$, which is therefore the locus of the
ring singularity. In particular, the vacuum solution of this class ($\m=0$, leading to $b^2=m^2/4-1$)
is singular. In the opposite case $m<2$, both brackets are positive definite, so that the
ring singularity is absent from the stationary region (it is hidden behind the horizon), all real values
of the rotation parameter $b$ being allowed from (\ref{parasB}). Note that in this case the ratio
$|J|/M^2=|b|/m$ is unbounded and can be
arbitrarily large, while the ratio $|\mu|/|J$ exceeds the Barrow-Gibbons \cite{BG} bound $1$.

The horizon angular velocity
 \be\lb{OmH1}
\Omega_H = \frac{4b}{\kappa[(m+2)^2+4b^2]}
 \ee
has the same sign as that of the total angular momentum. It is bounded above (in absolute magnitude) by
 \be
|\Omega_{\rm max}| = \frac1{\kappa(m+2)},
 \ee
value attained for $|b|=1+m/2$, and goes to zero both in the
static limit $b\to 0$ and in the limit $b\to\infty$. Note
that in this last limit $\omega-\Omega_H^{-1}\to0$ everywhere (not
only on the horizon). The conical singularity parameter $\alpha_H$ takes the value
 \be\lb{alphaHB}
\alpha_H = \frac{16[(m+2)^2+4b^2]}
{(m+2)^4[(m-2)^2+4b^2]}.
 \ee
For $b=0$ (Bonnor), $\alpha_H= 16/(m^2-4)^2>1$ (negative string tension) in
the range $m<2$ where the ring singularity is absent \cite{emparan}. When
the rotation parameter $b$ is turned on, $\alpha_H$ decreases until
a critical value $b=b_c$:
 \be\lb{bc}
b_c^2 = \frac{m(m+2)^2(8-m^2)}{4(m+4)(m^2+4m+8)}
 \ee
such that $\alpha_H(m,b_c(m))=1$, i.e. the conical
singularity is absent. $b_c$ is clearly real in
the range $0<m<2$. For $b>b_c$, $\alpha_H$ continues to decrease towards a limiting value
$16/(m+2)^4$ for $|b|\to\infty$.

It is generally assumed that, except in the special case of the static Majumdar-Papapetrou
linear superpositions,
multi-black hole systems are unbalanced, and the force necessary to stabilize such systems
is proportional to the tension $(1-\alpha_S)/4$ (or $(1-\hat\alpha_S)/4$ in the horizon co-rotating frame)
of the interconnecting cosmic string(s). Indeed it has been shown
\cite{DiHo85,SHR07} that balance between two non-degenerate black holes cannot be achieved without
intermediate conical singularities. The situation concerning rotating multi-black hole systems with
degenerate horizons is less clear. It is noteworthy that there is at least a two-parameter
subclass of Manko et al. solutions for which the conical singularity is absent, suggesting that these
systems may be in equilibrium. We will return to this point in the concluding section.

The values of the other horizon parameters are
 \ba\lb{horparaB}
M_H &=& \frac{4\kappa mb^2[(m-2)^2(m^2+2m+4) + 4b^2(m^2-2m+4)]}{(m+2)^2[(m-2)^2+4b^2]^2}, \nn\\
N_H &=& \frac{\kappa\alpha_Hm(m+2)^2b}8, \quad
P_H = \frac{\kappa\m(4-m^2-4b^2)}{(m+2)[(m-2)^2+4b^2]}, \nn\\
Q_H &=& \frac{8\kappa\m b}{(m+2)[(m-2)^2+4b^2]}, \quad \phi_H =
\frac{\kappa\m}m\Omega_H
 \ea
(the value of the horizon angular momentum $J_H$ can be retrieved from the horizon Smarr formula
(\ref{smarr})). The horizon Komar mass is positive definite.

The limiting case \underline{$b=0$} with $m<2$ corresponds to the Bonnor solution.
Eq. (\ref{horparaB}) shows that $M_H=0$ in this case. The fact that the horizon Komar mass of the Bonnor
magnetostatic solution vanishes, which does not seem to have been pointed out previously, is a direct
consequence of the Smarr relation for a static, electrically neutral field configuration with degenerate
horizons. It then follows from (\ref{totalmass}) that the mass of the Bonnor dihole must be equal to the
string Komar mass. This is checked in the Appendix by an independent direct evaluation of (\ref{MS}). The
only non-zero horizon observables of the Bonnor dihole are their magnetic charges $\pm P_H$, with
 \be
P_H = \frac{\kappa\m}{2-m},
 \ee
the dipole moment $2\kappa P_H$ accounting partly for the total magnetic moment $\mu = \kappa^2\m$, with
which it coincides in the limit $m\to0$.

\subsection{Large distance limit}
The large distance limit can be defined as the limit when the distance $2\kappa$ between the two horizons becomes
very large, $\kappa\to\infty$, while the total mass $M=\kappa m$ is held fixed, i.e. $m\to0$. In this limit
the ratio $|J|/M^2=|b|/m$ can be arbitrary large, unless $b$ goes also to zero. The horizon angular velocity and
the string tension
 \be
\Omega_H \sim \frac{b}{\kappa(1+b^2)}, \qquad \frac{1-\alpha_H}4 \sim \frac{Mb^2}{2\kappa(1+b^2)}
 \ee
both go to zero as $1/\kappa$. An exception is the Bonnor static dihole $b=0$, for which $\Omega_H=0$ but
$(1-\alpha_H)/4 \sim -M^2/8\kappa^2$ \cite{emparan}.

The large distance values of the other horizon observables are
 \ba
M_H &\sim& \frac{Mb^2}{1+b^2}, \quad N_H \sim \frac{Mb}2, \nn\\
Q_H &\sim& \varepsilon\frac{Mb}{\sqrt{1+b^2}}, \quad P_H \sim \varepsilon\frac{M(1-b^2)}{2\sqrt{1+b^2}}
 \ea
($\varepsilon=\pm1$), and\footnote{To evaluate the limit of $J_H$ one must expand $M_H$, $Q_H$ and $\phi_H$
to order O($m^2$), as the terms of order O($m$) cancel in (\ref{smarr}).}
 \be
J_H \sim \frac{M^2b(1-b^2)}{2(1+b^2)} \sim Q_HP_H.
 \ee
This last relation shows that in this limit the dyonic black hole angular
momentum $J_H$ is not an inertial effect (the horizon angular velocity goes to zero), but is purely electromagnetic.
Note also that for fixed $b$ the total angular momentum
 \be
J = \kappa Mb \sim 2\kappa N_H \gg J_H
 \ee
is due essentially to the NUT dipole.
If the parameter $b$ also goes to zero (slowly rotating Bonnor dihole), then as in the static Bonnor case all the
horizon observables go to zero, except for the horizon magnetic charge $P_H \sim \varepsilon M/2$, leading to the
magnetic dipole $\mu \sim 2\kappa P_H$.

An interesting special case is $|b|=1$. Then,
 \be
M_H \sim |N_H| \sim \frac{M}2, \quad J_H \sim 0, \quad |Q_H| \sim \frac{M}{\sqrt2}, \quad P_H \sim 0.
 \ee
In this case the string mass $M_S = M-2M_H$ goes to zero, while the horizon observables satisfy the static extremality condition
 \be
M_H^2+N_H^2 \sim \frac{M^2}2 \sim Q_H^2+P_H^2.
 \ee

\setcounter{equation}{0}
\section{The static class of solutions}
The horizon angular velocity vanishes, so that the constituent black holes become static, for
 \be
\lambda = \nu \quad \Leftrightarrow \quad b = - \frac{4\nu\delta}{m^2}
 \ee
which leads to
 \be
\delta = 1 + \nu^2 - \frac{m^2}4 \ge 0,
 \ee
this last restriction following from the condition
 \be
\m^2 = \frac{16\delta(\nu^2+d^2)}{m^2} \ge 0.
 \ee
If further $\nu=0$ we recover again the Bonnor magnetostatic solution
$a=b=0$ ($\lambda=\nu=0$, $d=m^2/4$).
Another special case is $\delta=0$ ($\nu^2=m^2/4-1$), leading to
the rotating TS2 vacuum solution \cite{TS}.

We shall refer to this class of solutions with non-rotating constituent black holes as the ``static''
class. Despite this appellation, the solutions are generically non-static, with net total angular
momentum $J=\kappa^2 ma$, where
 \be\lb{arotB}
a = \frac{2\nu(m^2-2\delta)}{m^2}.
 \ee
This angular momentum is due partly to the dipole moment of the two opposite NUT charges (\ref{N})
carried by the black hole horizons, and partly to the string angular momentum. The two exactly
balance, so that the rotation parameter (\ref{arotB}) vanishes, $a=0$, if $\delta = m^2/2$, corresponding to
 \be
4\nu^2 = 3m^2-4, \quad \m^2 = 2m^2(m^2-1) \qquad (m > 2/\sqrt3).
 \ee
This stationary magnetized solution, with non-rotating horizons and vanishing total angular momentum,
reduces for $\nu=0$ to the static Bonnor solution with the special value $m=2/\sqrt3$.

For the static class,
 \be
{\rm Re}(U+W) \equiv \zeta(\zeta+mx+2d)+ \nu[mbx-\nu(1+y^2)](1-y^2).
 \ee
While for the Bonnor solution this is well-known to be positive definite for $|m|<2$,
in the general case there does not seem to be a parameter range where ${\rm Re}(U+W)$ is
positive definite. For $y=0$ and $x=1$, one obtains
 \be\lb{ringRB}
{\rm Re}(U+W)(1,0) \equiv \frac{m+4}m\left[-\delta^2 + \delta + \frac{m(4-m^2)}{4(m+4)}\right].
 \ee
The discriminant of the trinomial in $\delta$ inside brackets
$$\Delta=\frac{(m+1)(-m^2+m+4)}{m+4}$$
is positive for $m<m_0=(1+\sqrt{17})/2\simeq2.56$, in which case (\ref{ringRB}) has two roots
$\delta_\pm=(1\pm\sqrt\Delta)/2$. So, given that ${\rm Re}(U+W)$ is positive for large $x$,
a necessary condition for the absence of equatorial ring singularity is
 \be\lb{regstat}
m < m_0, \quad \delta_- < \delta < \delta_+.
 \ee
The TS2 solution ($\delta=0<\delta_-$) is well-known to present a ring singularity.
For the static subclass $a=0$, the necessary condition ${\rm Re}(U+W)(1,0)>0$ reduces to
$$(m+1)(-m^3-3m^2+4m+4)>0,$$
leading to
 \be\lb{regstat0}
2/\sqrt3 \simeq 1.15 \le m < m_1 \simeq 1.49 \qquad (a=0).
 \ee

Because the horizon angular velocity vanishes, the conical singularity parameter is simply
$\alpha_H = {\hat\alpha}_S = \alpha_S = 1/(\delta^2+\nu^2)$. For the $a=0$ subclass, this gives
 \be
\alpha_H = \frac4{(m^2-1)(m^2+4)} \qquad (a=0).
 \ee
This is bounded above by $\alpha_{\rm max}= \alpha(2/\sqrt3) = 9/4$, and goes to zero for $m\to\infty$. So there is a
critical value
 \be\lb{critstat0}
m_c = [(-3+\sqrt{41})/2]^{1/2} \simeq 1.30 \qquad (a=0)
 \ee
such that the conical singularity is absent,
$\alpha_H(m_c)=1$. This value of $m_c$ satisfies the necessary regularity condition (\ref{regstat0}).

Although, strictly speaking, the Tomimatsu formulas we have used to compute the various horizon parameters
break down for $\Omega_H=0$, we can nevertheless use here our results (\ref{QH1}) and (\ref{MH1}), together
with (\ref{P}) and (\ref{N}), in the {\em limit} $\Omega_H\to0$ ($\lambda\to\nu$). We give here the results
only for the subclass \underline{$a=0$}:
 \ba\lb{parastat0}
N_H &=& - \frac{2\kappa\nu m(m+1)}{(m-1)(m^2+4)}, \quad P_H = - \frac{\kappa\m[2m^3-3m^2+4m-4]}{2(m-1)^2(m^2+4)}, \nn\\
Q_H &=& - -\frac{\kappa\m\nu(2-m)}{(m-1)^2(m^2+4)}, \quad \phi_H = - \frac{\m\nu}{2m(m^2-1)} \qquad (a=0).
 \ea
The horizon mass is given by (\ref{MH1}), which can be written as $M_H= M_g + M_e$, where the terms $M_g$ and $M_e$
coming from the two terms in the (modified) Tomimatsu formula (\ref{MH}) can be thought of, loosely speaking, as
the ``gravitational'' and ``electromagnetic'' contributions to the horizon Komar-Tomimatsu mass. For consistency,
we must check the Smarr formula (\ref{smarr}), which for $\Omega_H=0$ reduces to $M_H=\phi_HQ_H$. We obtain for $a=0$:
 \be
M_g = \frac{\kappa m^2(2-m)^2}{2(m-1)(m^2+4)}, \quad M_e = \frac{\kappa m(2-m)[2m^3-3m^2+4m-4]}{4(m-1)^2(m^2+4)},
 \ee
leading to
 \be
M_H =  \frac{\kappa m(2-m)(3m^2-4)}{4(m-1)^2(m^2+4)} = \phi_HQ_H \qquad (a=0).
 \ee
This is positive definite in the regularity range (\ref{regstat0}). One can check that the string mass
$M_S=M-2M_H$ is also positive in this range.

\setcounter{equation}{0}
\section{Summary and discussion}
We have analyzed the four-parameter class of asymptotically flat magnetized solutions to the Einstein-Maxwell 
equations constructed
in \cite{manko00a}, and shown that these represent systems of two co-rotating extreme black holes with equal
masses and electric charges, and opposite magnetic and NUT charges, connected by a cosmic string.
A special two-parameter subclass of solutions, without ring singularity, was previously discussed in
\cite{tale}. We have discussed here in some detail several other three-parameter subclasses, and determined
in each case the parameter domain in which the ring singularity is absent. For the ``neutral'' class
(electrically neutral constituents), there are two regularity sectors. In the first sector the string tension
is positive, while it is negative in the second sector. In both sectors, the horizon Komar mass is
negative. For the Bonnor class, there is only one regularity sector, with varying string tension,
which vanishes in a two-parameter subclass. The horizon Komar mass is always positive, except for the static
Bonnor solution itself, where it vanishes. We have also considered for this class the large distance limit
$\kappa\to\infty$, $m\to0$. Finally, for the ``static'' class (rotating solutions with non-rotating
constituents), there is also a single regularity sector. We have discussed in more detail the two-parameter
subclass such that the total angular momentum vanishes, and have found that for this subclass the horizon
mass, as well as the string mass, are positive. Again, the string tension vanishes in a one-parameter family
of this subclass.

A remarkable finding is the existence of rotating systems of extreme black holes with vanishing tension of
the interconnecting string. This calls for two observations. In the presence of NUT charges, this string
is also a Misner string, with spin $\hat\omega_S$ (in the horizon co-rotating frame) proportional
to the NUT charge, see (\ref{nut}). If the horizons are rotating, then the string tension in the global
frame (that of an observer at spatial infinity) is different from the string tension in the local horizon
co-rotating frame, as follows from the last equation (\ref{stringpara}). So our statement that the string
tension vanishes is ambiguous. Actually, it is valid in the horizon co-rotating frame, where
$\hat\alpha_S=\alpha_H=1$ means that the constituent black holes feel no tension, so that their horizon is
smooth. Yet the asymptotic observer can in principle measure a tension $(1-\alpha_S)/4$, with
 \be
\alpha_S = \frac{\alpha_H}{1-4N_H\Omega_H}.
 \ee
For the Bonnor subclass $(\kappa,m,b_c(m))$ (with $b_c$ given by (\ref{bc})), the horizon angular
velocity and NUT charge are of the same sign (see (\ref{OmH1}) and (\ref{horparaB})), so that $\alpha_S>1$,
corresponding to a negative observed string tension, to compensate an observed attraction between
the two black holes. On the other hand, our statement that there is a one-parameter (the scale $\kappa$)
family of solutions in the subclass $a=0$ of the static class without a conical singularity is unambiguous,
as the horizons are non-rotating.

The second observation is that for axisymmetric solutions of Einstein's equations, written in the Weyl
form
 \be\lb{weyl}
ds^2 = -F(dt-\omega d\varphi)^2 +
F^{-1}[e^{2k}(d\rho^2+dz^2)+\rho^2d\varphi^2],
 \ee
there are two regularity conditions on the symmetry axis $\rho=0$, $k(0,z)=0$, and $\omega(0,z)=0$.
These are satisfied by construction on the two semi-infinite portions $y=\pm1$, $x>1$ of the axis,
but not, generically, on the interconnecting string $x=1$, $-1<y<1$, where they translate into $\alpha_S=1$,
$\omega_S=0$, or $\hat\alpha_S=1$ and $\hat\omega_S=0$ in the horizon co-rotating frame. In other words, 
in the unambiguous `static' case with vanishing string tension, there remains a Misner string singularity.
We expect that, contrary to a cosmic string, where geodesics terminate, this Misner string will be
transparent to geodesic motion, as shown in \cite{GC15} in the cases of the Taub-NUT metric and of the 
dyonic Reissner-Nordstr\"om-NUT metric. However there is always, in the vicinity of a Misner string, an 
`acausal' region containing closed timelike curves (CTC). We have argued in \cite{GC15,GC18} that the 
existence of such a region where $g_{\varphi\varphi}=0$ does not necessarily lead to observable violations 
of causality. While this matter clearly deserves further investigation, we feel that the possibility of 
spacetimes with finite-length Misner strings and the associated compact CTC regions should be left open.

The preliminary analysis reported here should be extended to a more systematic investigation of this
four-parameter class of solutions, starting with the determination in the general case of the 
parameter domain for which the ring singularity is absent.

\section*{Acknowledgments}
I warmly thank Dmitry Gal'tsov for a critical reading of the manuscript and useful suggestions.

\renewcommand{\theequation}{A.\arabic{equation}}
\setcounter{equation}{0}
\section*{Appendix: Computation of the Bonnor string mass}
This is given by (\ref{MS}), which reads in the case of the Bonnor magnetostatic solution
 \be\lb{MSB}
M_S = {\rm lim}_{(\xi\to0)}\,\frac14\int_{-1}^{+1}\left[g^{\xi\xi}g^{tt}\partial_\xi g_{tt}
- 2A_\varphi F^{\xi\varphi})\right]\sqrt{|g|}dy,
 \ee
with $\xi^2\equiv{x^2-1}$. The first, purely gravitational term does not contribute in the limit
$\xi\to0$ because $g_{tt} = {\rm constant} + {\rm O}(\xi^2)$.  From (\ref{ansatz}) and (\ref{sol})
with $b=\nu=0$, we obtain
 \be
A_\varphi = \frac{\kappa\m(1-y^2)(2x+m)}{2[x^2-1+\delta(1-y^2)]},
 \ee
leading to
 \be
F_{\xi\varphi} = - \frac{\kappa\m(\tau^2+y^2)}{\delta(1-y^2)}\,\xi + {\rm O}(\xi^3),
 \ee
where we have put
$$ \tau^2 \equiv \frac{m+2-\delta}\delta = \frac{m+2}{2-m}. $$
Using the string metric (\ref{string}) for $b=\nu=0$, where
$$\Sigma_S = \delta^4(1-y^2)^2(\tau^2-y^2)^2 + {\rm O}(\xi^2),$$
we then obtain on the string $\xi=0$
 \be
A_\varphi = \frac{\kappa\m(m+2)}{2\delta}, \quad \sqrt{|g|}F^{\xi\varphi} = - \frac\m\delta\,\frac{\tau^2+y^2}{(\tau^2-y^2)^2},
 \ee
leading to
 \be
M_S = \frac{2\kappa m^2}{2-m}\,\frac1{\tau^2-1} = \kappa m = M.
 \ee


\begin{thebibliography}{9}

\bb{bachweyl} R. Bach and H. Weyl, Math. Zeits. {\bf 13} (1922) 134.

\bb{israelkhan} W. Israel and K.A. Khan, Nuovo Cimento {\bf 33} (1964) 331.

\bibitem{papa} A. Papapetrou, Proc. Roy. Irish Acad. A \textbf{51} (1947) 191;
S.D. Majumdar, Phys. Rev. \textbf{72} (1947) 390.

\bb{weinstein96} G. Weinstein, Commun. Part. Diff. Eq. {\bf 21} (1996) 1389
[arXiv:gr-qc/0001081].

\bb{GP} J.B. Griffiths and J. Podolsky, Exact space-times in Einstein's General Relativity, CUP, 2009.

\bb{2kerr} I. Cabrera-Munguia, V.E. Ceron, L.A. L\'opez and Omar Pedraza, Phys. Lett. B {\bf 772} (2017) 10
[arXiv:1702.02209]; V.S. Manko and E. Ruiz, Phys. Rev. D {\bf 96} (2017) 104016 [arXiv:1702.05802].

\bb{emparan} R. Emparan, Phys. Rev. D {\bf61} (2000) 104009 [arXiv:hep-th/9906160].

\bb{bonnor66} W.B. Bonnor, Z. Phys. {\bf 190} (1966) 444.

\bb{tale} G. Cl\'ement and D. Gal'tsov, Phys. Lett. B {\bf 771} (2017) 457 [arXiv:1705.08017];
``Stationary binary black holes without naked ring singularity'' [arXiv:1806.11193].

\bb{GC98} G. Cl\'ement, Phys. Rev. D {\bf 57} (1998) 4885
[arXiv:gr-qc/9710109].

\bb{sibga84} N.R. Sibgatullin: Oscillations and Waves in Strong Gravitational
and Electromagnetic Fields (Nauka, Moscow, 1984;
English translation: Springer-Verlag, Berlin, 1991).

\bibitem{manko00a} V.S.~Manko, E.W. Mielke and J.D.~Sanabria-Gomez,
Phys.\ Rev.\ D {\bf 61} (2000) 081501 [arXiv:gr-qc/0001081].

\bibitem{manko00b} V.S.~Manko, J.D.~Sanabria-Gomez and O.V.~Manko,
Phys.\ Rev.\ D {\bf 62} (2000) 044048.

\bb{smarr} G. Cl\'ement and D. Gal'tsov, Phys. Lett. B {\bf 773} (2017) 290
[arXiv:1707.01332].

\bibitem{ZV} D.M. Zipoy, J.\ Math.\ Phys.  {\bf 7} (1966) 1137; 
B.~H.~Voorhees, Phys.\ Rev.\ D {\bf 2} (1970) 2119.

\bb{TS} A. Tomimatsu and H. Sato, Phys. Rev. Lett. {\bf 29} (1972) 1344;
Progr. Theor. Phys. {\bf 50} (1973) 95.

\bb{KoHi} H. Kodama and W. Hikida, Class. Quantum Grav. {\bf 20},
5121 (2003) [arXiv: gr-qc/0304064].

\bb{tom84} A. Tomimatsu, Progr. Theor. Phys. {\bf 72} (1984) 73.

\bb{tom83} A. Tomimatsu, Prog. Theor. Phys. {\bf 70} (1983) 385.

\bb{BG} J.D. Barrow and G.W. Gibbons, Phys. Rev. D {\bf 95} (2017) 064040
[arXiv:1701.06343].

\bb{DiHo85} W. Dietz and C. Hoenselaers, Ann. Phys. {\bf 165} (1985) 319.

\bb{SHR07} J. Sod-Hoffs and E.D. Rodchenko, Class. Quantum Grav. {\bf 24}
(2007) 4617 [arXiv:0705.3973].

\bibitem{GC15}
G.~Cl\'ement, D.~Gal'tsov and M.~Guenouche, Phys. Lett. B {\bf 750} (2015)
591 [arXiv:1508.07622]; Phys. Rev. D {\bf 93} (2016) 024048 
[arXiv:1509.07854].

\bb{GC18} G. Cl\'ement and M. Guenouche, Gen. Rel. Grav. {\bf 50} (2018) 60
[arXiv:1606.08457].

\end{thebibliography}
\end{document}